# Optical properties of atomically thin transition metal dichalcogenides: Observations and puzzles

M. Koperski,[†,§] M. R. Molas,[†] A. Arora,[†,‡] K. Nogajewski,[†] A. O. Slobodeniuk,[†] C. Faugeras,[†] M. Potemski[†]

[†]Laboratoire National des Champs Magnétiques Intenses, CNRS-UGA-UPS-INSA-EMFL, 25 rue des Martyrs, 38042 Grenoble, France

[§]Institute of Experimental Physics, Faculty of Physics, University of Warsaw, Pasteura 5, 02-093 Warszawa, Poland

[‡]Institute of Physics and Center for Nanotechnology, University of Münster, Wilhelm-Klemm-Strasse 10, 48149 Münster, Germany

**Abstract**

Recent results on the optical properties of mono- and few-layers of semiconducting transition metal dichalcogenides are reviewed. Experimental observations are presented and discussed in the frame of existing models, highlighting the limits of our understanding in this emerging field of research. We first introduce the representative band structure of these systems and their interband optical transitions. The effect of an external magnetic field is then considered to discuss Zeeman spectroscopy and optical pumping experiments, both revealing phenomena related to the valley degree of freedom. Finally, we discuss the observation of single photon emitters in different types of layered materials, including wide band gap hexagonal boron nitride. While going through these topics, we try to focus on open questions and on experimental observations, which do not yet have a clear explanation.

## 1. Introduction

Atomically thin layers of semiconducting transition metal dichalcogenides (S-TMDs) represent a new class of materials which are of vivid interest [1–6], primary in the area of the semiconductor physics and nanoscience as well as opto-electronic applications. The studies of thin films of S-TMDs are greatly inspired by and profit from the research and developments focused on graphene-based systems [7]. Viewed as a part of the "beyond graphene research", the studies of thin films of S-TMDs are largely at the level of basic research, aiming to enlighten their fundamental properties and potentially new functionalities [1–6]. The investigations of optical properties represent one of the main directions of research on S-TMD layers. Selected topics in this area are presented here, in particular those covered to a certain extent by a series of our recent studies [8–14]. Considerable emphases are focused on striking effects and unresolved problems which are discussed in the context of the apparent experimental data and related theoretical concepts.

## 2. Overview of electronic bands; optically bright and darkish monolayers

The family of S-TMDs (of $MX_2$ type) includes $MoS_2$, $MoSe_2$, $WS_2$, $WSe_2$, and $MoTe_2$ compounds. A large portion of this paper is focused on the properties of S-TMD monolayers (1L). 1L S-TMDs emit light efficiently [5] and are most commonly believed to be direct band gap two-dimensional semiconductors [5], in strike contrast to indirect band gap S-TMDs in the bulk form, extensively investigated in the past [15]. A pictorial scheme of the near band-edge structure of 1L S-TMD is shown in Fig. 1. Minima (maxima) of the conduction (valence) band are located at the $K^+$ and $K^-$ points of the Brillouin zone (BZ) of the hexagonal crystal structure of S-TMDs. Within a simplified approach, this band structure follows the model of graphene with a broken sublattice symmetry (alternating M and X atoms in the lattice hexagons) [16–19], which implies an appearance of the gap as well as the specific, valley selective rules for circular polarization of optical transitions (see Fig. 1). The next relevant characteristics of S-TMDs' bands are the significant effects of the spin orbit (SO) interaction, inherited from heavy transition metal (M) atoms [16]. The states around the valence band (VB) maxima are at large built of M-orbitals with an angular momentum 2 ($d_{\pm 2}$), thus displaying rather large SO splitting ($\Delta_{so,vb}$), which indeed varies from ~150 meV for $MoS_2$ to ~450 meV for $WSe_2$ [8,10,11,13,20–35], see Table 1. $\Delta_{so,vb}$ accounts, in the first approximation, for the energy separation between two most characteristic interband transitions observed in S-TMDs known as A and B optical resonances (see Fig. 1). Since the states around the conduction band (CB) minima are predominantly built of $d_0$ M-orbitals, they show a considerably smaller SO splitting ($\Delta_{so,cb}$). Nevertheless, $\Delta_{so,cb}$ remains to be apparent due to the second order effects for $d_0$ orbitals and an admixture of $p$-states of chalcogen (X) atoms [32,34,36]. Notably, these two latter effects are expected to yield the opposite in sign contributions to the SO splitting, thus raising a possibility of two distinct scenarios of: (i) optically bright monolayers with the aligned spins in the upper VB and lowest CB subbands ($\Delta_{so,cb}> 0$, set as a convention) and thus optically active ground state transition, and ii) optically darkish monolayers with antiparallel spins in the upper VB and lowest CB subbands ($\Delta_{so,cb}< 0$), thus optically inactive ground state transition. The band structure calculations indicate that $MoSe_2$ and $MoTe_2$ monolayers can be bright whereas $WSe_2$ and $WS_2$ monolayers are likely darkish [32,34,36], see Table 1. The case of $MoS_2$ is less clear; the estimated $\Delta_{so,cb}$ for this material falls in the range of small meVs. When comparing the optical response of different monolayers we argue in this paper that their bright or darkish character is one of the important ingredients to be accounted for their often quite distinct and sometimes conspicuous behavior. The quantitatively different dependence of the photoluminescence (PL) intensity versus temperature, observed in $WSe_2$ [10,37–39] and $MoSe_2$ [11,37] monolayers has been the first experimental signature of possibly distinct band alignments in these two monolayers. Typically, as shown in Fig. 2, the PL intensity decays considerably with increasing the temperature in 1L $MoSe_2$ [11]. However, an unconventional rise of the PL signal can be observed in 1L $WSe_2$ [10]. Different hypotheses have to be thoroughly examined to firmly account for these experimental observations, but one of them follows our assertion of distinct alignments of SO split subbands in 1L $MoSe_2$ and 1L $WSe_2$. Indeed, if the energetically lowest lying interband resonance is optically inactive in 1L $WSe_2$, the increase of the PL intensity with temperature in this monolayer is accounted for by a considerable activation of the population of higher energy, optically active states. In contrast, the $MoSe_2$ monolayer with optically active ground state resonance shows a conventional decrease of the PL intensity with increasing the temperature reflecting, as for many semiconductors, a gain in efficiency of non-radiative processes at high temperatures.

Whereas S-TMD monolayers are commonly considered to be direct gap semiconductors, most of N-monolayers (NLs) of S-TMDs emit light much less efficiently and all of them appear to be indirect gap semiconductors for N > 1 (perhaps with an exception of a possible direct band gap $MoTe_2$ bilayer [8]). This crossover from direct to indirect band gap system is reproduced in a number of band structure calculations [2,17–19]. As illustrated in Fig. 1, the relevant band structure evolution expected upon increasing N consists of the upwards shift of the valence band at the Γ point of the BZ and the downwards shift of the conduction band at the Λ point (located roughly in the middle way of the band dispersion in the direction between K and Γ points of the BZ, see Fig. 1). As a result of these band shifts, the N-MLs with N > 1 appear as semiconductors with indirect Λ-Γ or K-Γ band gaps. Worth noting is, however, that the pronounced absorption band edge (e.g., that traced in absorption-type experiments) is always (in all NL S-TMDs) associated with the direct band gap appearing at the $K^+$ and $K^-$ points of the BZ [2,17–19].

Single-particle band structure picture is certainly not sufficient to accurately describe the optical properties of S-TMD layers. Coulomb interaction between electrons and holes is particularly strong in these systems, imposing the exciton binding energies in the range of few hundreds of meVs in monolayers [28,29,33,40–45]. The unconventional dielectric screening effects for 2D S-TMDs deposited on alien substrates is another relevant issue (e.g., resulting in non-Rydberg ladders of excitonic excited states in 1L S-TMDs [11,28,42]). Here we bear in mind the excitonic effects at a simplified but commonly referred level, assuming the formation of bound exciton states associated with each separate pair of the VB and CB subband edges.

Firm description of the optical properties of S-TMDs requires the band structure and excitonic effects to be taken into account on the same footing. Such models are, however, largely not analytical and so far mostly applied to a single system of $MoS_2$ monolayer [40,41,46]. Whereas the works on $MoS_2$ structures have indeed pioneered the research on atomically thin S-TMDs, the optical quality (width of resonances) of $MoS_2$ films remains rather poor, even today (see [47] for efforts to improve the quality of $MoS_2$ optical response). Considerably better quality of optical responses (narrower resonances) is displayed by other S-TMD layers and those are mainly discussed in the course of this paper.

## 3. Fundamental interband optical transitions

The fundamental optical transitions in S-TMD monolayers, appearing in the vicinity of A excitons, span the spectral range from ~1.15 eV for 1L $MoTe_2$ [8,48] up to ~2.1 eV for 1L $WS_2$ [5,23,49]. These transitions are most commonly traced with photoluminescence and reflectance/absorption measurements. The results of such measurements (at low temperatures of about 10 K) are shown in Fig. 3 for four representative S-TMD monolayers of $MoTe_2$, $MoSe_2$, $WSe_2$ and $WS_2$ (intentionally undoped flakes deposited on $Si/SiO_2$ substrates).

To start with we focus on the **absorption response of S-TMD monolayers** (red traces in Fig. 3), as derived from transfer-matrix analysis [50,51] of the measured reflectance contrast spectra (light-grey curves). Please note that direct comparison of the measured reflectance-contrast spectra is rather meaningless due to different dielectric/interference effects in multilayer geometry of S-TMD flakes deposited on $Si/SiO_2$ substrates with, in particular, different thicknesses of $SiO_2$ layers. As largely discussed in literature, one expects the fundamental absorption spectra of S-TMD monolayers to display strong Lorentzian-shape resonances due to neutral A exciton ($X_A^0$), and possibly, lower-energy, usually weaker peaks, commonly assigned to charged excitons (trions) $X^\pm$ [52] (due to unintentional, n- or p-type doping, likely present in majority of S-TMD samples). As can be seen in Fig. 3, the above scenario holds for the case of 1L $MoTe_2$ [13] and $WSe_2$, but not for $MoSe_2$ and $WS_2$ [53]. Strikingly, the shape of the $X_A^0$ absorption in 1L $MoSe_2$ (marked as 1 in Fig. 3(A)) deviates strongly from the expected Lorentzian shape and appears more to be a dispersive-like [11].

The polariton- [54] or Fano-type coupling [55] are among known effects which can account for a dispersive shape of absorption resonances. The polariton effects should be excluded in our structures without any optical cavity which is otherwise required to observe the effects of coupling of 2D excitons with photon modes [56,57]. On the other hand, the Fano-type resonances may arise when an excitonic state interacts with the continuum of states of another lower-energy resonance, as has been previously observed in III-V semiconductors and their heterostructures [58–61], and also for the so called M-point excitons in graphene [62,63]. In reference to our observation, we speculate that the continuum of states spanned by the charged-exciton coincides in energy with the ground state exciton, and a strong interaction can exist [64,65]. This hypothesis has been further supported by the reflectivity-contrast measurements performed as a function of temperature [11], see Fig. 3(B). It can be noticed that as the temperature is increased, the charged-exciton gradually loses its absorption cross-section and almost completely disappears for $T > 200$ K. This is accompanied by a drastic change in the Fano-like line shape, which transforms into an expected Lorentzian-like for $T > 200$ K. The Fano-type effect due to the coupling of the $X_A^0$ state with a continuum of the $X^\pm$ state is a possible explanation of the dispersive absorption shape in 1L MoSe$_2$. This hypothesis calls, however, to be verified as it raises some pertinent questions, such as, for example, what is specific to the MoSe$_2$ monolayer with respect to all other monolayers which behave more conventionally and do not show the Fano-type resonances.

A simple picture of the appearance of two ($X_A^0$ and $X^\pm$) resonances is also not consistent with what is seen in the fundamental absorption of WS$_2$ monolayers [53]. As is often observed in our experiments, an additional, third line (see feature marked 2 in Fig. 3(A)) appears in the 1L WS$_2$ spectra, on the low energy side of the $X^\pm$ resonance. At this moment, we can only speculate that this additional resonance might be due to weak absorption associated with the impurity bound excitons [9,10,20,29,66–78] and/or reflecting the appearance of optically active $X^\pm$ transitions in both singlet and triplet configurations with different binding energies [77,79,80]. Both scenarios are feasible, but more work is needed to uncover the origin of the unusual triple-peak absorption observed in 1L WS$_2$. Again, the question about the particularity of one (this time WS$_2$) among others S-TMD monolayers remains to be clarified.

Focusing now on low-temperature **photoluminescence response of S-TMD monolayers** one immediately notices (see Fig. 3) a quantitative difference between the PL spectra characteristic of MoSe$_2$ and MoTe$_2$ monolayers and those displayed by WSe$_2$ and WS$_2$ monolayers. Whereas the PL spectra of MoSe$_2$ and MoTe$_2$ monolayers are rather simple and mostly show just two well-defined peaks, the low-temperature PL response of WSe$_2$ and WS$_2$ monolayers is by far more complex (multitude of emission peaks).

The two PL peaks in MoSe$_2$ and MoTe$_2$ monolayers are associated with radiative recombination of the neutral ($X_A^0$) and charged ($X^\pm$) A excitons [8,11,13]. Notably, these two PL transitions find their counterparts in the observed absorption/reflectance resonances, appearing at similar energies. The relative intensity of the $X_A^0$ and $X^\pm$ resonances, observed in the PL and absorption-type spectra, depends on many factors; the density of excess carriers being one of the crucial [52]. Less pronounced $X^\pm$ resonances in our 1L MoTe$_2$ than in 1L MoSe$_2$ may reflect the effect of smaller in the former case (but higher in 1L MoSe$_2$) unintentional doping of the sample.

The two ($X_A^0$ and $X^\pm$) resonances can be also identified in the PL spectra of WSe$_2$ and WS$_2$ monolayers (as coinciding in energy with their counterparts in absorption response) but these spectra are in fact dominated by the bands of emission lines (highlighted by dashed circles in Fig. 3), which appear on the low energy side of the charged exciton's line. These dominant below-exciton emission bands, reported in many papers [9,10,20,29,66–78], are commonly but vaguely assigned to radiative recombination of localized-exciton complexes, with excitons bound to defects/crystal imperfections/impurities [81]. Disorder is certainly an important factor when interpreting the PL spectra of semiconductors. However, it is very surprising that the "exciton localization" effects are so much pronounced in WSe$_2$ and WS$_2$ monolayers but barely apparent in 1L MoTe$_2$ and MoSe$_2$. Thus, it is tempting to speculate that the apearance of the below-exciton band in S-TMD monolayers is largely a subject of whether a given monolayer is bright (optically active ground state exciton, $\Delta_{so,cb} > 0$) or darkish (optically inactive spin-forbiden ground state exciton, $\Delta_{so,cb} < 0$). In the PL process at low temperature, the photocreated carriers largely populate the ground state excitons which then efficiently recombine (directly or with the assistance of a charge carrier) in case of bright monolayers. Instead, the dark ground state excitons in WSe$_2$ and WS$_2$ monolayers cannot recombine directly but require the assistance of disorder and/or phonons to break the spin or k-vector selection rules [82,83]. The below-exciton emission band observed in S-TMD monolayers might therefore well be due to recombination of dark-exciton states, though activated by disorder or phonons [82,83]. Further studies are needed to clarify the above point and, in particular, the case of MoS$_2$ should be more examined with respect to its bright or darkish character. It should be noted that the pronounced below-exciton band can be observed for MoS$_2$ monolayers as well [20,66,67,78]. Following our reasoning, the MoS$_2$ monolayer should be classified as darkish, which, however, is in contradiction with a theoretical estimation of its very small but still positive $\Delta_{so,cb} \sim 3$ meV. The latter value refers to the amplitude of the SO splitting at the edge of the conduction band and may change the sign for only slightly higher k-values [32]. A firm assessment of the character of the ground-state-exciton state in 1L MoS$_2$ remains perhaps an open question.

Well-pronounced **A and B excitons** also dominate the absorption onset **in S-TMD multilayers (NL)**, in spite of the fact that all NL S-TMDs with N > 1 are indirect band gap semiconductors (see Fig. 1). The evolution of the low-temperature absorption spectra (as extracted from the reflectance-contrast measurements) as a function of number of layers is shown in Fig. 4 for WSe$_2$ and MoSe$_2$ systems. The distinct effect visible in these spectra, which we would like to stress and discuss here, is a clear experimental trend that the energy positions of the most pronounced A and B excitonic resonances are largely independent on N. This might be surprising since one expects that the band structure of NL S-TMDs changes profoundly with N [2,19]. In addition, the strength of the electron-hole Coulomb binding is also strongly dependent on layer number, for instance, the A-exciton binding energy is estimated to be in the range of 50 - 90 meV in bulk S-TMDs [10,11,15,84,85] but reaches the values close to 0.5 eV in S-TMD monolayers [28,42–44,86]. The observation that the energies of A and B excitons in NL S-TMDs are practically independent on N implies that the expected shrinkage of the bands at the K points of the BZ is practically fully compensated by the decrease in the exciton's binding energy, see Fig. 5. This result is somewhat reproduced in advanced although not analytical calculations [46], though one would perhaps expect more intuitive arguments to account for this effect. It is worth noticing that the robustness of excitons (with respect to their apparent energy position) is also known under different conditions, e.g. for S-TMD monolayers embedded in different dielectric environment [87–89]. The unchanging energy positions of the A and B excitons in multilayers and in monolayers with diferent dielectric enviroment may have a similar background. Even in multilayers, the A and B excitons (wave functions) are well confined in each individual layer [88], thus their modification (as a function of number of layers) may be thought of as mostly determined by the changes of the surrounding dielectric response (as in the case of monolayers on different substrates). To this end, similar compensation effects are also known to appear in conventional two-dimensional systems in the presence of excess charged carriers (see e.g., [90]) and/or for high density electron-hole systems (see e.g., [91]). Each time the band gap renormalization effects are concluded to be compensated by the decrease of the exciton binding energy. Whether all these "compensation effects" may or may not have a common explanation is perhaps another pertinent question.

## 4. Zeeman spectroscopy of excitonic resonances in S-TMD monolayers

Investigation of excitonic resonances in the presence of an external magnetic field is a commonly exploited approach to get a deeper insight into the electronic properties of semiconducting materials. Recently, multiple experiments in the magneto-optical domain have been reported for S-TMD monolayers [13,92–97]. The primary effect reported in these works and further discussed here is the linear with the magnetic field B (applied in the direction across the layer), Zeeman-type splitting of optical transitions into two components observed in the $\sigma^+$ and $\sigma^-$ configurations of circularly polarized light. This Zeeman splitting is conveniently parametrized in terms of the effective g-factor which accounts for the amplitude and sign of the energy difference between the $\sigma^+$ and $\sigma^-$ polarized components of a given optical transition [6,67,98]: $\Delta E = E(\sigma^+) - E(\sigma^-) = g\,\mu_B\,B$, where $\mu_B \approx 0.058$ meV/T stands for the Bohr magneton. The effective g-factors of interband optical transitions are commonly assumed to reflect the relative g-factors characteristic of the respective conduction and valence band edges (at the K$^+$ and K$^-$ points of the BZ in 1L S-TMDs), thus providing a trustworthy test of the band structure models [95]. It is worth to mention that all possible contributions to the Zeeman effects in 1L S-TMDs (i.e., due to spin, atomic orbitals and band carriers' motion) are not easy to be compiled into a single self-consistent theoretical model [6,95].

The understanding of Zeeman effects in S-TMD monolayers is rather poor. The observed excitonic g-factors remain controversial as they are not soundly accounted for by the band structure models which, in the ultimate case, imply even practically vanishing excitonic Zeeman effect, in strike contrast to the experimental data [95] (notably, the fact that the excitonic wave functions in 1L S-TMDs are spatially very localized [28,42] and therefore built of a broad spectrum of states in the momentum space, seems to be not relevant to weaken this controversy). The extremely large g-factors (up to $g \approx -15$) of certain PL lines observed in 1L WSe$_2$ (and WS$_2$) [9] are by far surprising. Whereas the magneto-optical data indicate a rather large Zeeman effect for the upper valence subbands of S-TMD monolayers, the recent magneto-transport results on p-doped WSe$_2$ monolayers [99], may, at very first sight, be seen as implying this Zeeman splitting to be rather small (inferior to the Landau level separation due to orbital effects). Here we present an attempt to elaborate, on the phenomenological level, a conceivable picture of linear in B shifts of the band edges in S-TMD monolayers, which accounts for the main experimental observations and enlighten the apparent controversy on Zeeman effects in these systems.

The fundamental optical resonances in 1L S-TMDs are, within a simple model, associated with the interband transitions involving the edges of overall eight conduction and valence subbands, four degenerated at B = 0 pairs of subbands, see Fig. 6. To test the Zeeman effect for these subbands, we first note that simple time reversal symmetry arguments imply that: $\Delta E_{c(v),\uparrow(\downarrow)}^{K^+}(B) = -\Delta E_{c(v),\downarrow(\uparrow)}^{K^-}(B)$, where $\Delta E(B)$ is the linear with B energy shift of a conduction (c) or valence (v) band state with the spin up ($\uparrow$) or spin down ($\downarrow$) orientation at the K$^+$ or K$^-$ points of the BZ. Following the above set of four trivial conditions, the Zeeman effect for our eight electronic subbands can be parametrized with four parameters, the number of which can be, however, further reduced by taking into account the prominent experimental observations. One of the robust experimental results follows the comparison of the Zeeman patterns for the A and B excitonic resonances, observed in a number of magneto-absorption/reflectance measurements of 1L S-TMDs [13,97,100]. An example of such a measurement is shown in Fig. 6. It illustrates a typical effect of the applied magnetic field on the reflectance-contrast spectrum, represented here for a MoSe$_2$ monolayer measured at low temperatures. The A and B excitonic resonances yield the pronounced spectral features; each of them splits into two $\sigma^+$ and $\sigma^-$ polarized components when the magnetic field is applied perpendicular to the layer plane, in the Faraday geometry (see the spectra measured at B=30T). Noteworthy is a weaker feature on the low energy side of the A resonance, which is assigned to the charged exciton resonance [24] and, as expected, observed only in one polarization configuration ($\sigma^+$) at high magnetic fields., i.e., when all excess free carriers are fully spin/valley polarized. Returning to the main A and B excitonic resonances, we observe that the energy separation between their $\sigma^+$ and $\sigma^-$ components is linear with the strength of the magnetic field and according to our convention this excitonic Zeeman effect is characterized by negative g-factors ($\sigma^-$ components appear at higher energies than their $\sigma^+$ counterparts). Most important for our further analysis is the experimental fact that both A and B resonances display practically identical g-factors. As summarized in Table 2, this experimental rule applies to most of S-TMD monolayers (the case of 1L MoTe$_2$ is a possible exception). When accepting the rule of identical g-factors for A and B excitons in a given 1L S-TMD we impose an additional condition on the Zeeman shifts of our conduction and valence band subbands: $E_{c,\uparrow}^{K\pm}(B) - E_{v,\uparrow}^{K\pm}(B) = E_{c,\downarrow}^{K\pm}(B) - E_{v,\downarrow}^{K\pm}(B)$. Taking this condition into account, the Zeeman shifts of the eight subbands can be described with just three parameters which, generally speaking, can be chosen in different ways, but one of them has a rather transparent physical interpretation. Following the latter choice, we express the Zeeman shifts of all eight CB and VB subbands in terms of parameters: $E_{d_2}$, $E_S$, and $E_V$ as:

$$\begin{cases} \Delta E_{c,\uparrow}^{K\pm}(B) = \pm E_V + E_S \\ \Delta E_{c,\downarrow}^{K\pm}(B) = \pm E_V - E_S \\ \Delta E_{v,\uparrow}^{K\pm}(B) = \pm E_V + E_S \pm E_{d_2} \\ \Delta E_{v,\downarrow}^{K\pm}(B) = \pm E_V - E_S \pm E_{d_2} \end{cases}$$

The $E_{d_2}$ term appears only for the VB subbands imposing their valley-selective Zeeman shifts, $\pm E_{d_2}$ for the $K^{\pm}$ subbands. The $E_S$ term is spin-selective and results in $+E_S$, $-E_S$ Zeeman shifts for the spin up and the spin down subband, respectively. Finally, the valley term shifts all (CB and VB) $K^+$ subbands by the same value $E_V$ and all $K^-$ subbands by $-E_S$. The above phenomenological parametrization stems from general knowledge [93,96] that the Zeeman shifts for fundamental CB and VB subbands of a 1L S-TMD might be composed of three possible contributions: due to the orbital effects at the atomic level ($E_{d_2}$, indeed expected for VB states built of d$_2$ orbitals and not for CB states built of d$_0$ orbitals), due to the genuine electronic spins ($E_S$), and due to possible different orbital motion of carriers in the $K^+$ and $K^-$ bands with the opposite chiralities (valley term $E_V$). Arguments have been put forward [93] that all three parameters: $E_{d_2}$, $E_S$, and $E_V$ should be positive and such an assumption is kept in the following analysis. A pictorial representation of the way they alter the energy of states of a S-TMD monolayer in a magnetic field is given in Fig. 7. We believe that the scheme of Zeeman effects follows the same set of equations for all S-TMD monolayers. It is then important to note that, as depicted in Fig. 7, the scheme of Zeeman effects with respect to the lower and upper CB subbands is different for the two types of previously introduced, bright and darkish monolayers [10,11,38].

In the attempt to estimate the amplitude of the Zeeman effects in S-TMD monolayers (i.e., to estimate the amplitude of each $E_{d_2}$, $E_S$, $E_V$ terms), we first assume that the spin term for the electrons in S-TMD monolayers is the same as for a free electron in vacuum [93], namely: $E_S = 1\mu_B B$. Next, we refer to the magneto-optical studies and note (see Fig. 7) that the Zeeman splitting of any optically active interband transition within our eight CB and CV subbands is expected to be solely given by the orbital term $E_{d_2}$ (the $E_S$ and $E_V$ terms cancel each other for optically allowed transitions which conserve the spin as well as the valley index/pseudospin). The $E_{d_2}$ term can be therefore estimated from the magneto-reflectance measurements of the A and B exciton resonances as $g_{X_A^0}\mu_B B = g_{X_A^0}\mu_B B = -2E_{d_2}$ where $g_{X_A^0} \approx -4$ and $g_{X_A^0} \approx -4$ stand for the approximate amplitude of the g-factors, correspondingly for the A and B excitons, most commonly reported in the literature (see

[93,95,96] and Table 2). To estimate the third missing parameter we refer to the recent magneto-transport experiments reported for p-type WSe$_2$ monolayers [99]. These experiments have been performed on relatively highly doped samples, though still with the Fermi energy located within the upper valence band subbands, and in the range of magnetic fields when many Landau levels remain populated (Landau level filling factors, v > 4). The measured Shubnikov de Haas (SdH) oscillations at low magnetic fields, consistent with the Hall voltage measurements, show the persistence of doubly degenerated Landau levels at low magnetic fields and lifting of this degeneracy only in the limit of high magnetic fields. As already mentioned, this result, reminding a typical observations for a 2DEG in GaAs structure, could, at a very first sight, point out a rather small Zeeman splitting ($\Delta E_Z$) for holes in 1L WSe$_2$, i.e., smaller than the separation between Landau levels (cyclotron energy, $\tilde{\hbar}\omega_C$). The reported SdH oscillations [99] probe, however, LLs with high indices and the conclusions drawn from the magneto-transport data may also be valid under different conditions, i.e., when $\Delta E_Z \approx N \tilde{\hbar}\omega_C$ where N is not only zero but could be a small even integer (N = 2 or N = 4) as well. In view of our previous discussion, $\Delta E_Z = 2\left(E_S + E_{d_2} + E_V\right) \approx 2(\mu_B B + 2\mu_B B + E_V)$ is rather big, thus, if the $E_V$ term is not anomalously large, one should expect that the condition $\Delta E_Z \approx 2 \tilde{\hbar}\omega_C$ holds for holes in 1L WSe$_2$. Finally, we derive $\Delta E_Z \approx 9 \mu_B B$ and $E_V \approx 1.5\mu_B B$, using the effective mass of holes $m^* = 0.45m_e$ and come up with the final estimation of $E_v \approx 1.5\mu_B B$, $E_s \approx 1\mu_B B$ and $E_{d_2} \approx 2\mu_B B$ for a 1L WSe$_2$ but believe that pretty similar values of these parameters can be applied to other 1L TMDs as well.

The investigations of the magneto-PL spectra provide an additional test for the proposed scheme of Zeeman effects in S-TMD monolayers. These spectra, we believe, are markedly different for bright and darkish S-TMD monolayers, in contrast to quantitatively similar Zeeman patterns of excitonic resonances in these two types of systems, observed in absorption/reflectance measurements. Fig. 8 illustrates the $\sigma^+/\sigma^-$ polarization-resolved magneto-PL spectra of two representative MoSe$_2$ and WSe$_2$ monolayers. The MoSe$_2$ monolayer, expected to be the optically bright system, exhibits a quite conventional magnetic field evolution of its PL spectra. The two observed emission lines due to recombination of the neutral and charged exciton split in magnetic field yielding the g-factors equal to about −4, similarly to their counterparts observed in magneto-reflectance spectra. As already discussed in the previous sections, the PL spectra of the WSe$_2$ monolayer are much more complex and moreover show an intriguing evolution with the magnetic field. Although the neutral exciton ($X_A^0$) emission line can be well recognized in these spectra and undergoes the expected Zeeman splitting ($g \approx -4$), there is also a multitude of emission peaks which appear on the low energy side of the $X_A^0$ line and display quite striking evolution with the magnetic field: the g-factors associated with these emission peaks vary from $g = -4$ up to $g = -15$. As already mentioned, we speculate that numerous emission peaks appearing in optically darkish systems, such as the WSe$_2$ monolayer, may be due to recombination of dark excitons, allowed by disorder or phonon-assisted processes. As illustrated in Fig. 9, the imaginable recombination processes which violate the k-vector or spin selection rules would, in accordance with the proposed scheme of the Zeeman effect in S-TMD monolayers, result in anomalously large g-factors for the emission lines in darkish systems.

To summarize this section, we conclude that the application of magnetic field represents a relevant test for the band structure models of S-TMD materials and that the magneto-optical data are in support of our conjecture of the appearance of two distinct, optically bright and darkish S-TMD monolayers. Obviously, more works are needed to confirm our conclusions on a more quantitative level.

## 5. Optical orientation of valley pseudospin in S-TMD monolayers

Interband excitation of a semiconductor brings in a possibility to transfer the angular momentum of circularly polarized photons to photoexcited carriers, to create a non-equilibrium orientation of their spins and, eventually, to examine the conservation of this orientation in the crystal by probing the polarization degree $P$ of the emitted light. When the excitation light is $\sigma^+$ ($\sigma^-$) polarized, $P$ is defined as:

$$P = \frac{I(\sigma^{+(-)}) - I(\sigma^{-(+)})}{I(\sigma^+) + I(\sigma^-)}$$

where $I(\sigma^{+(-)})$ denotes the intensity of the $\sigma^+$ ($\sigma^-$) polarized component of the emitted light (PL signal). Optical orientation studies have been widely explored in zinc-blende semiconductors with respect to the angular momentum of electronic spins [101,102]. Such studies are of vivid interest in monolayers of S-TMDs [20,67,68,98,103–106], in which the circular polarization of light ($\sigma^\pm$) is coupled to the valley degree of freedom (pseudospin) ($K^\pm$). The expected robustness of the valley pseudospin [66], together with the possibility of its optical orientation at room temperature [107,108] are promising for designing opto-valleytronic devices.

Two examples of the results of optical orientation experiments performed, at liquid helium temperature, on monolayers of WSe$_2$ (left panel) and of WS$_2$ (right panel) are shown in Fig. 10. As discussed in the previous sections, the spectra of these monolayers are rather complex and apart from the common features due to neutral ($X_{A0}^0$) and charged ($X^\pm$) excitons they display also a series of intense emission lines at lower energies, which are commonly assigned in literature to localized excitons and are here speculated to imply the recombination of dark excitons.

The polarization degree $P$ is presented in both panels of Fig. 10, as a black curve. $P$ displays a clear dependence on the emission energy and can be as large as 50% in the high energy part of the PL spectra. Notably, the polarization degree is non-zero also in the energy range of the so-called localized exciton emission. The results presented in Fig. 10, as well as many other reported previously in literature [20,67], show that the photo-excited S-TMD monolayers (WSe$_2$, MoS$_2$, as well as MoS$_2$) conserve the information on polarization, and that the population of photo-excited carriers in a given valley can be efficiently initialized. Such experiments established S-TMDs as a possible platform for opto-valleytronic devices and led to the observation of valley coherence [98,109]. Importantly, the optical orientation in some, e.g., MoS$_2$ [20,67], S-TMD monolayers remains efficient even at room temperature. It is worth noting that the degree of polarization also strongly depends on the excitation laser energy and a significant decrease in the efficiency of the optical orientation has been observed when increasing the excitation laser energy away from the emission energy range [107]. Exciting carriers at higher energies may induce significant losses in the polarization degree either due to less efficient injection of valley-polarized carriers or due to more efficient inter-valley scattering during the carrier relaxation processes.

Whereas more or less efficient optical orientation is easily observed for WSe$_2$, WS$_2$ and MoS$_2$ monolayers [20,67,68], the polarization memory effects are practically absent in MoSe$_2$ and MoTe$_2$ monolayers [110,111]. This is somewhat surprising as the overall band structure of all these monolayers is pretty similar. Elucidating this problem is of relevant interest for our understanding of the electronic properties of S-TMD materials. It should find its explanation either in the details of the band structure and/or in a qualitatively distinct for different monolayers, efficiency of the inter-valley scattering processes [82].

The hypothesis which we put forward here is that the optical orientation is effective in darkish but not efficient in bright S-TMD monolayers (following the arguments invoked in previous sections, we classify the 1L MoS$_2$ as a darkish system). As discussed below, a number of experimental observations can be explained in this frame, notably, when taking into account the efficient channel of the disorientation of the valley pseudospin in S-TMD monolayers, which is due to the strong, electron-hole exchange interaction [82,112–115] within the optically active excitonic states [32,82]. In bright monolayers the optically active excitonic states are the ground states, thus they are most populated (long lived) under the optical excitation, in favor of the efficient intervalley scattering. In contrast, the lifetime of optically active excitons in darkish monolayers can be rather short (as they are the excited exciton states in these systems), i.e., shorter that the inter-valley relaxation time, thus accounting for the efficient optical orientation in these systems. If the above qualitative arguments hold, the bright monolayers should also exhibit the optical orientation effects, but likely on a very short time scale. The recombination process in S-TMD monolayers are known to be fast [107], and it is only recently that the study of the dynamics of the photo-excited carriers in S-TMD monolayers has become available on short, sub-picosecond time scale, by implementing the four-wave mixing micro-spectroscopy [14,109]. Such experiments have been performed on $MoSe_2$ monolayers and indeed show the non-zero initial degree of polarization in this bright monolayer but at the same time a rapid loss of the polarization memory on ~1 ps time scale, see Fig. 11. In favor of our hypothesis is also the observation of a significant decrease of the polarization degree when using an excitation resonant with the low-energy exciton of a 1L $WS_2$. This effect is presented in Fig. 12, which illustrates the evolution of the degree of polarization in 1L $WS_2$ when sweeping the excitation laser energy through the neutral exciton energy. A pronounced decrease of the degree of polarization is observed when exciting resonantly into this optically active excitonic state. This observation is in line with the efficient depolarization mechanism acting at the level of the bright exciton. The resonant excitation into bright exciton states likely enhances their population with respect to lower energy dark states, thus enhancing the efficiency of inter valley scattering.

Finally, we note that the degree of optical orientation in $WSe_2$ monolayers has been recently reported to be significantly increased by the application of tiny magnetic fields [12]. This effect is illustrated Fig. 13. The application of as small magnetic field as 100 mT across the 2D layer is indeed sufficient to enhance the polarization degree of the PL spectra by a factor of more than two. The measured photoluminescence spectra in the absence and in the presence of magnetic field are shown in the left panel of Fig. 13, and the polarization degree as a function of the magnetic field, in the mT range is displayed in the right panel of this figure. Similar effects are observed for $WS_2$, but not for $MoSe_2$ monolayers [116]. Arguments have been put forward [12] that tuning the efficiency of optical orientation is possible in darkish but not in bright S-TMD monolayers. The application of a small magnetic field can only alter the weak inter-valley scattering process within dark excitons states because of the expected large Zeeman splitting of these states.

To conclude this section, we believe to have presented convincing experimental arguments that optical orientation of the valley pseudospin is effective in darkish monolayers but it is much less efficient in bright S-TMD monolayers.

## 6. Single photon emitters in S-TMDs and h-BN structures

In addition to a variety of intriguing and not yet fully understood properties discussed so far, the optical response of selected thin-layer members of the S-TMD family has been recently shown to exhibit narrow emission lines. These lines, with the ultimate full width at half maximum, FWHM, smaller than 100 μeV, appear at the selected locations on S-TMD flakes, in the energy range below the PL signal from 2D excitons in monolayer films [9,69–72,117–124], which rise much broader, a few meVs wide, emission peaks. The most extensive experimental work has been done till now in this respect on $WSe_2$ [9,69–72,117–123]. Much less abundant but equally convincing data has been presented also for $WS_2$ [121,122]. In the case of $MoSe_2$ [124], some key findings are still missing to firmly conclude that narrow emission lines observed on this material are of exactly the same type as in WSe$_2$ and WS$_2$, even though all the results obtained up to now support such a statement. For MoS$_2$ and MoTe$_2$, the literature lacks any reports demonstrating the existence of narrow emission lines in their optical response.

The narrow emission lines observed on WSe$_2$ have been unequivocally proven to originate from centers possessing a characteristic attribute of single-photon sources. Indeed, these PL lines show a prominent photon antibunching in the second-order autocorrelation function $g_2(\tau)$ at zero delay time ($\tau$) between photon counts in the two arms of the Hanbury Brown and Twiss interferometer. An example of such a result is illustrated in Fig. 14. It shows a low-temperature PL spectrum of a selected narrow-line-emitting center (NLEC) located at the edge of about 8 nm thick WSe$_2$ flake in the upper panel and the photon autocorrelation measurements done on one of the most intense narrow lines in that spectrum in the bottom panel. All NLECs reported in the literature for WSe$_2$ display similar lifetimes of the emitting state, which vary from a few hundreds of picoseconds up to a few nanoseconds as deduced from the coincidence time for photon antibunching and independently confirmed by time-resolved PL measurements [70,71]. An enhancement of the excited state's lifetime by one to two orders of magnitude with respect to the low-temperature lifetime of free excitons in monolayer WSe$_2$ is indicative of the 3D quantum confinement [72]. The quantum nature of NLECs found in this material is also highlighted by a characteristic saturation of their emission intensity at high excitation power, similar to that known from the physics of two-level systems [70,72,117]. The typical emission energies of NLECs fall in the range of 20 to 200 meV below the PL of delocalized 2D excitons in monolayer WSe$_2$, and overlap with a broad emission band usually ascribed to bound/localized excitons. This suggests that optical properties of NLECs are inherently linked to those of the 2D monolayer. Further support for such a statement is provided by photoluminescence excitation data [9] and the results of polarization-resolved magneto-PL measurements [9,70–72]. The latter studies unveil a similar, anomalously large Zeeman effect (g-factor amplitudes ranging from ~8 to ~13) for both the narrow emission lines and the 2D localize/bound excitons.

The origin and structural characteristics of the NLECs in WSe$_2$ still remain a matter of debate. In samples prepared by means of mechanical exfoliation of bulk crystals, the NLECs occur either at the flake's edges (including interfaces between parts of composite flakes) [9,69,71,72] or in the close vicinity of structural defects in the form of wrinkles, folds, or bubbles of air trapped between the flake and the substrate [117]. The former possibility is not restricted to flakes of any particular thickness, while the latter one applies only to flakes composed of one or two WSe$_2$ layers [117,121]. In the case of monolayers grown by chemical methods [70], only the air bubbles caused by locally weak adhesion of the flake to the substrate play an important role for the NLECs' formation. For each of the two groups of crystal imperfections described above, the appearance of the resulting NLECs seems to be different. The edge- and interface-related NLECs are rather expected to have the form of nano-sized pieces of WSe$_2$ monolayer as the one shown in Fig. 15, with STM (scanning tunneling microscopy) measurements done on the perimeter of a few-layer-thick WSe$_2$ flake transferred on top of graphene grown on SiC substrate. Instead, the NLECs associated with wrinkles, folds, and air bubbles are considered to originate from strain-induced localization of 2D excitons. In either case the NLECs can be thought of as quantum dots (QDs) with quite shallow confinement barriers of about 1.5-2.5 meV [9,72] as deduced from temperature evolution of the narrow line emissions which typically disappear from the PL spectrum above 20-30 K. The way the narrow emission lines broaden with increasing the temperature resembles the acoustic phonon broadening observed for single QDs in conventional semiconductors [9]. The QD-like picture of the NLECs in WSe$_2$ is also supported by zero-field splitting (0.4-1 meV) of their emission lines into two components of nearly orthogonal linear polarizations and comparable intensities [70–72,117–120]. This is similar to the fine structure splitting of the neutral exciton emission in conventional QDs (which is due to the QD shape anisotropy and the electron-hole exchange interaction). Last but not least, the QD scenario for the NLECs sounds plausible because of very recent demonstration of their capability to emit single photons in a sequential manner due to the biexciton-exciton recombination cascade [120]. This prominent result is presented in Fig. 16. It shows a pair of polarization-resolved PL spectra comprising two doublets of narrow emission lines that come from the recombination of a biexciton (lines P1 and P3) and exciton (lines P2 and P4), respectively, as well as the photon cross-correlation measurements performed between the lines P1 and P2 under both pulsed (main graph in panel (b)) and continuous-wave (inset to the main graph) excitation. A clear bunching effect observed at zero delay time in the histogram of photons emitted under pulsed excitation and a characteristic transition from an anti-bunching for negative delay times to bunching at small positive delay times seen in the continuous-wave measurements prove that the biexciton and exciton recombination form a cascade.

Despite all the experimental observations listed above, the NLECs in $WSe_2$ are far from being well-understood and well-controllable objects. They still require deeper structural characterization that would tell us more about their origin and properties. The possibilities of creating them on demand by scratching the host monolayer with a microscope tip [69] or by strain engineering [117,122,123], as well as of filling them with charge carriers not only by optical but also electrical means [118,119,121] look very promising from the point of view of their potential applications in future optoelectronic devices. First, however, the threshold temperature for disappearance of the NLECs has to be moved well beyond the present 20-30 K limit by elaborating the methods to deepen their localization potential.

As already mentioned, the available experimental data on about NLECs in $WS_2$ is not as abundant as for $WSe_2$. Even though locating such objects in the flakes prepared by means of mechanical exfoliation does not appear to be a particularly difficult task, no deeper analysis of NLECs in $WS_2$ based on PL spectra like the one shown in Fig. 17(A) has been published so far [122]. The emission of narrow lines in this material has been, however, observed in electro-luminescence (EL) experiments [121] and indeed shown to display the character of quantum emitters. The elaboration of their properties including in particular a structural characterization and a study as a function of the magnetic field and temperature still remains to be done.

Whereas NLECs in $WS_2$ were revealed with EL experiments, numerous facts about the QD-like emission occurring in the low-temperature PL spectra of monolayer $MoSe_2$ have been established so far [124]. Similarly to the case of $WSe_2$, the narrow emission lines in $MoSe_2$ (FWHM ranging from 150 to 400 μeV) appear at energies below the PL due to 2D excitons (neutral and charged). A majority of them does not display any measurable zero field splitting although it is possible to find NLECs whose emission lines are split at 0 T by as much as 0.8 meV. When applying the magnetic field , the emission lines due to NLECs exhibits the same Zeeman effect as the emission related to delocalized 2D excitons in the monolayer, with identical sign and similar amplitudes of the g-factors equal to about -4 in either case. Finally, in full agreement with $WSe_2$, the NLECs in monolayer $MoSe_2$ emerge in the vicinity of defects in the form of wrinkles or folds, which means that an important role in their formation is played by strain-induced potential wells. The assignment of their quantum-emitter character (with photon correlation measurements) is, however, still missing. Problems with acquiring such data most probably stem from the fact that, as shown in Fig. 17(B), unlike $WSe_2$ and $WS_2$, the narrow emission lines in the PL spectrum of $MoSe_2$ appear to be rather closely spaced what makes it difficult to select from them just a single line. A possible way to overcome this obstacle could be to deliberately apply a significant strain to the $MoSe_2$ flakes, within the spirit of such a method successfully employed to better select the NLECs in $WSe_2$ [117,122,123] and $WS_2$ [122].

Recently, a rich family of bright quantum emitters has also been discovered in hexagonal boron nitride (h-BN) [125–133], an insulating material with a band gap of about 6 eV that nowadays plays an important role in the field of graphene- and S-TMD-based van der Waals heterostructures, mostly serving as thin tunneling barriers in light-emitting diodes (see e.g., Refs [118], [119], and [121]) or as protective films for encapsulation of flakes that are particularly sensitive to environmental conditions like thin layers of $NbSe_2$ and other representatives of metallic subfamily of TMD compounds. The excitement about these single-photon sources that has emerged in the community of solid state physicists and engineers results from three main observations. First of all, they are capable of sustaining their quantum character (see Fig. 18) up to room temperature, which from the application point of view gives them a substantial advantage over the NLECs in S-TMDs. Secondly, their emission energy range covers both the visible [125,126,128–133] and UV [127] part of the electromagnetic spectrum. Finally, they seem to be present in all currently available forms of h-BN i.e. bulk single crystals, micrometer- and submicrometer-sized powder, as well as exfoliated and CVD-grown mono- and multilayers. Moreover, the NLECs in h-BN structures are characterized by high chemical resistance against such aggressive agents as hydrogen, oxygen, or ammonia and offer a possibility of creating them on demand by annealing, electron-beam irradiation, helium or nitrogen ion bombardment and wet-etching of host h-BN crystals [127,128,132].

The nature of NLECs in h-BN must be very much different from that of centers giving rise to single-photon emission in S-TMDs. Most importantly, they have nothing to do with recombination of band excitons as the energy range in which they emit and can be excited falls well below the band gap of h-BN. It means that similarly to colored centers in diamond, SiC or ZnO, the NLECs in h-BN originate from localized crystal structure imperfections like, e.g. nitrogen and boron vacancies, $N_BV_N$ defects created by nitrogen atoms, which move from their default sites to the neighboring boron sites and leave the former ones unoccupied, or carbon atoms incorporated into the crystal lattice at nitrogen sites during the growth process. There is no doubt that different narrow emission lines observed in the PL spectra of h-BN excited at below-band gap energies correspond to different structural imperfections. Their precise assignment to particular point defects is now the key challenge to be faced in order to deeply understand their origin and to be able to generate them in a controllable and reproducible way.

## 7. Conclusions

To summarize, we have reviewed a number of topics in the domain of optical and electronic properties of semiconducting transition metal dichalcogenides (S-TMDs). These topics follow a series of our recent studies which have been discussed in the context of a worldwide abounded research activity in the area of 2D TMD materials. Largely discussed here, and the most studied systems so far, are the S-TMD monolayers. Although all $MoS_2$, $MoSe_2$, $WSe_2$, $WS_2$, and $MoTe_2$ monolayers are known to be efficient light emitters and direct band gap 2D semiconductors, we believe that two quantitatively different classes of S-TMD monolayers, referred here to as bright and darkish, should be necessarily distinguished. These two type of monolayers display distinct alignment of the spin-orbit split subbands in the conduction band, and we argue that this fact is of particular importance when interpreting an often striking optical response of these systems. The role of non-radiative, e.g., Auger type processes, and/or kinetics of photo-excited carriers in TMDs is another issue to be explored [134,135]. In the course of our review, the considerable emphasis has been put on surprising observations which include those related to the appearance of quantum emitters in S-TMD structures and in another layered material - h-BN.

To this end we are convinced that research on thin layers of S-TMDs will be a blooming field of science in the future and that the applications of these systems will follow the pertinent research efforts. Among possible future studies are those more focused on gated structures, with a view to gaining better understanding of the optical (and magneto-optical) response of S-TMD monolayers in the presence of free carriers (with a controlled density). Another relevant direction are perhaps the investigations of monolayers under intense optical excitation and/or in optical cavities in search for the effects of interactions (e.g., exciton polaritons, many-body effects at high densities of electron-hole pairs) and for an efficient laser action in S-TMD materials. Largely unexplored so far are the S-TMD multilayers which may be interesting as well, for instance with respect to the observation of indirect excitons and their possible collective states. One could further speculate about many other possible directions in the research on S-TMD structures, but as it is often the case for a new and dynamically developing research field, these directions may be drastically changed with unexpected discoveries of completely new effects.

**Acknowledgements**

We thank D. Basko, V. I. Fal'ko, M. Orlita, P. Kossacki, J. Binder, T. Kazimierczuk, and T. Smoleński for helpful discussions. This work has been supported by the European Research Council (MOMB project no. 320590), the EC Graphene Flagship project (no. 604391), and the National Science Center (grant no. DEC-2013/10/M/ST3/00791), and has profited from the access to the Nanofab facility of the Institut Néel, CNRS UGA.

**Figures/tables**

**Tab. 1.** Amplitude of the spin-splitting at the K points of the Brillouin zone in the conduction ($\Delta_{so,cb}$) and valence ($\Delta_{so,vb}$) bands in monolayer S-TMDs deduced from theoretical predictions and experimental results [8,10,11,13,20–35]

|  | $MoS_2$ | $MoSe_2$ | $WS_2$ | $WSe_2$ | $MoTe_2$ |
|---|---|---|---|---|---|
| $\Delta_{so,cb}$ (meV) | 3 | 20 - 22 | -29 - -32 | -36 - -37 | 32 - 36 |
| $\Delta_{so,vb}$ (meV) | 138 - 150 | 180 - 202 | 379 - 429 | 400 - 510 | 213 - 269 |

**Tab. 2.** A comparison between the values of the g-factors of the neutral excitons A and B for monolayers of different representatives of S-TMDs, as obtained from the measurements of the reflectivity or transmission spectra. In most cases, within the experimental uncertainty the values are the same, which is used as a condition for derivation of the empirical model describing the magnetic field evolution of electronic states in S-TMD monolayers.

| material | neutral exciton A g-factor | neutral exciton B g-factor |
|---|---|---|
| $MoS_2$ | -4.0±0.2* *(reflectance)* | -4.2±0.2* *(reflectance)* |
| $MoSe_2$ | -4.2±0.2 *(reflectance)* | -4.2±0.2 *(reflectance)* |
| $MoTe_2$ | -4.8±0.2** *(reflectance)* | -3.8±0.2** *(reflectance)* |
| $WS_2$ | -4.3±0.2 *(transmission)* <br> -3.9±0.2* *(reflectance)* | -4.3±0.2 *(transmission)* <br> -4.0±0.2* *(reflectance)* |
| $WSe_2$ | -3.8±0.2 *(reflectance)* | -3.9±0.2 *(reflectance)* |

☐ – „darkish" materials   ☐ – bright materials

\* - data from Stier AV, *et al*. Nat. Comm. 2016, 7, 10643
\*\* - data from Arora A, *et al*. Nano Lett. 2016, 16, 3624

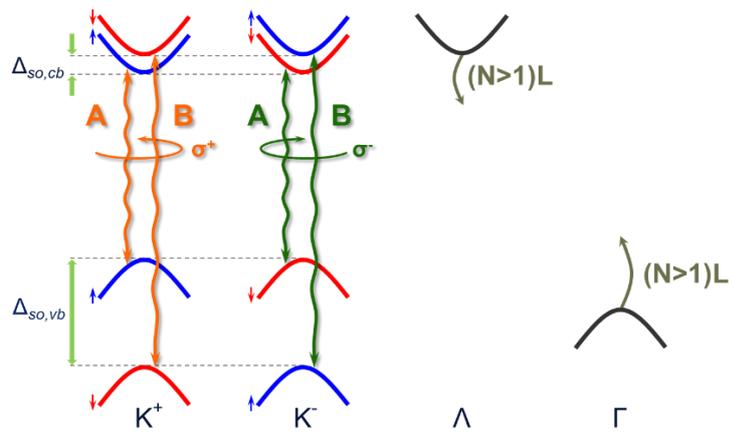

**Fig. 1.** Diagram of subbands in the conduction and valence bands at the $K^+$, $K^-$, $\Lambda$, and $\Gamma$ points of the Brillouin zone in monolayer $MoSe_2$. The blue (red) curves indicate the spin-up (spin-down) subbands, while the grey ones illustrate bands without the spin projection. The orange and green wavy lines show the A and B transitions associated with the $\sigma^+$ and $\sigma^-$ polarizations, respectively. $\Delta_{so,cb}$ and $\Delta_{so,vb}$ denote the corresponding spin-orbit splitting in the conduction and valence bands.

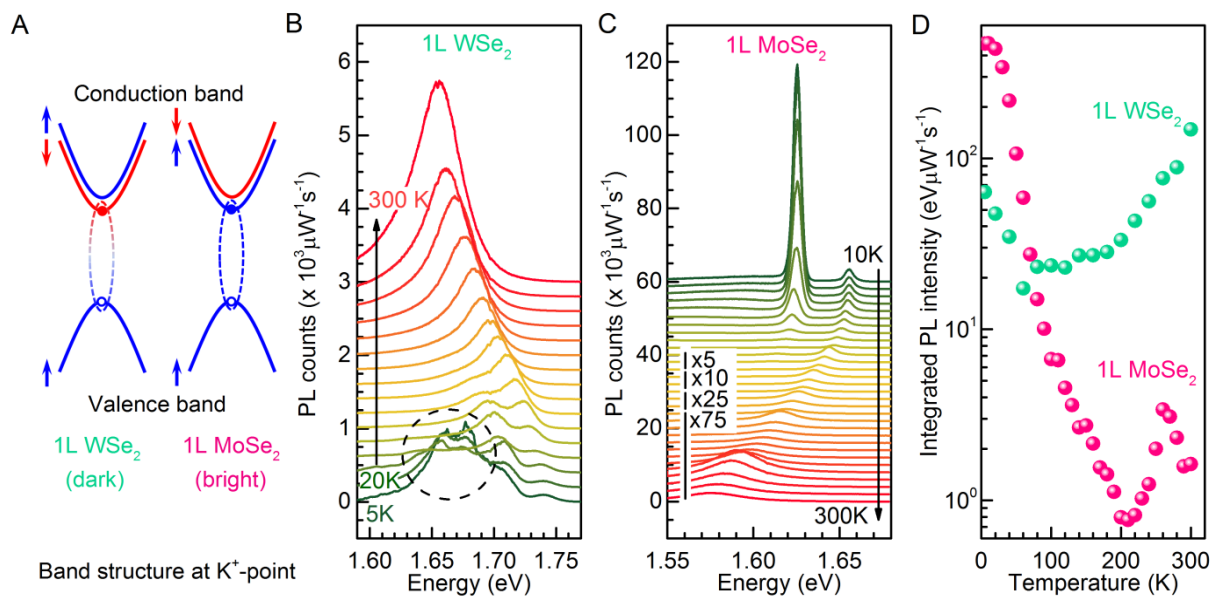

**Fig. 2. (A)** Valence band and spin-split conduction band at the $K^+$ point of the Brillouin zone for monolayer $WSe_2$ and $MoSe_2$. Due to the spin configuration of the carriers in the respective bands, the ground-state exciton is dark in $WSe_2$, whereas it is bright in $MoSe_2$. Photoluminescence spectra of **(B)** 1L $WSe_2$ [10], and **(C)** 1L $MoSe_2$ [11] as a function of temperature up to 300 K, in steps of 20 K and 10 K, respectively. The spectra have been shifted vertically for clarity, and are also amplified in **(C)** by factors mentioned in the figure. **(D)** Integrated PL intensity as a function of temperature depicting peculiar trends for the two materials [10,11].

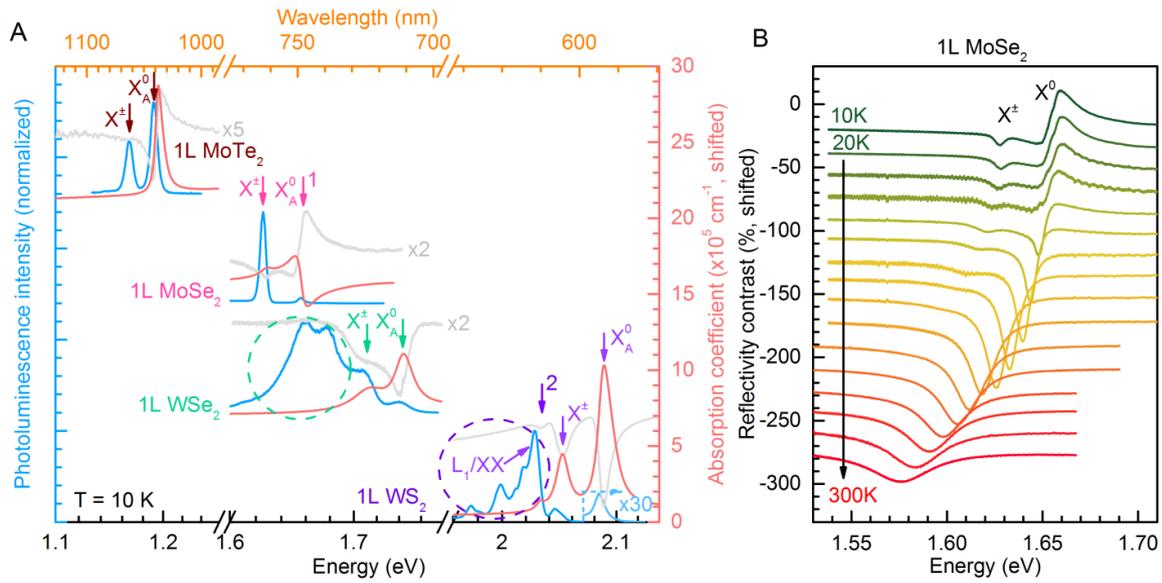

**Fig. 3. (A)** Photoluminescence (blue, normalized, left y-axis) and absorption spectra (red, right y-axis) of monolayers of WS$_2$ [53], WSe$_2$ [10], MoSe$_2$ [11], and MoTe$_2$ [13], in the energy region in the vicinity of A exciton. Absorption spectra have been derived from the transfer matrix based analysis of the measured reflectance-contrast spectra (light-grey, amplified with factors mentioned along-with). The neutral ($X_A^0$) and charged ($X^\pm$) A exciton transitions have been marked. The spectra for WSe$_2$, MoSe$_2$ and MoTe$_2$ have been shifted for clarity along the y-axis. The dashed circles highlight peculiar emission from WS$_2$ and WSe$_2$, while the transitions marked (1) and (2) represent tentatively assigned Fano-like and defect-related absorption peaks in monolayer MoSe$_2$ and WS$_2$, respectively. **(B)** Reflectance contrast spectra of 1L MoSe$_2$ as a function of temperature, in steps of 20 K, depicting the evolution of Fano-type resonance at low temperature, to a vanishing Fano-like effect at higher temperature [11].

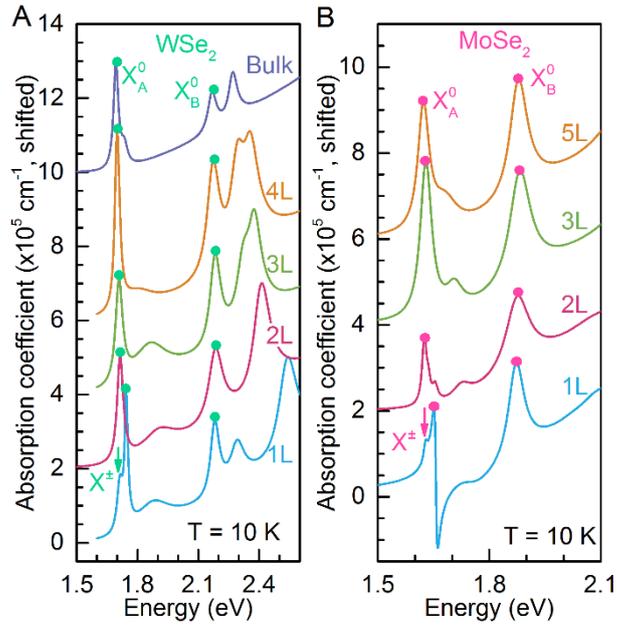

**Fig. 4.** Absorption spectra (derived from reflectance contrast spectra) of **(A)** WSe$_2$ [10] and **(B)** MoSe$_2$ [11] with increasing layer thickness, mentioned along with the respective spectra. The A and B exciton transitions, i.e. $X_A^0$ and $X_B^0$, are marked by green and pink filled circles for **(A)** and **(B)**, respectively. Additionally, a charged exciton's peak is highlighted for the 1L case in the two plots by arrows.

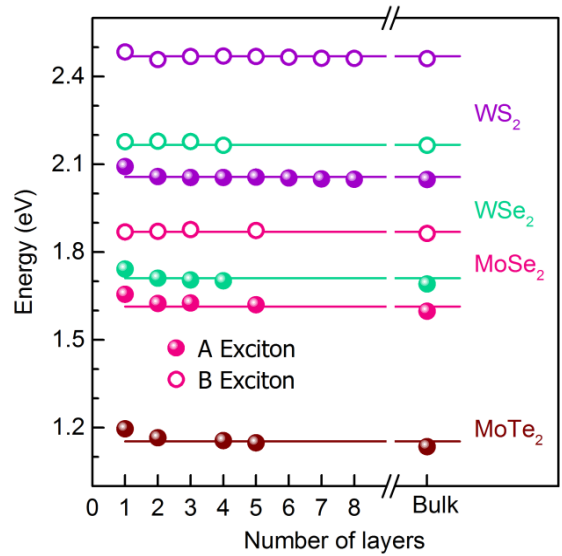

**Fig. 5.** Energies of A and B exciton resonances as a function of number of layers, in different S-TMD multilayers, as determined from the absorption type (reflectance contrast) measurements. The data for $WS_2$, $WSe_2$, $MoSe_2$, and $MoTe_2$ multilayers are purple-, green-, pink- and brown-coded, respectively. The results are compiled from the reports [8,10,11] and the recently obtained data (for $WS_2$) by M. R. Molas *et. al* [53]. The unpublished A exciton transition energy for bulk $MoTe_2$ is also shown.

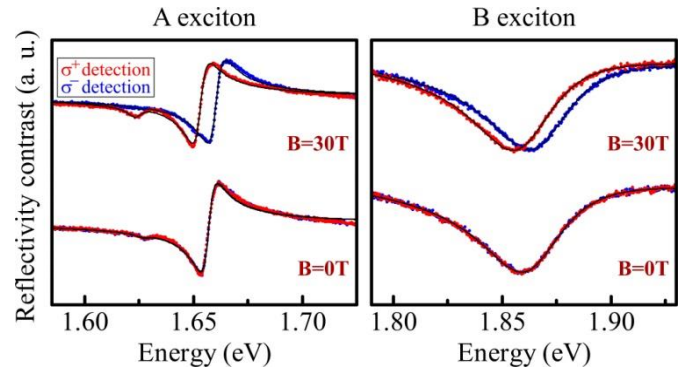

**Fig. 6.** Circular-polarization-resolved ($\sigma^+$ and $\sigma^-$ components) reflectance spectra of an MoSe$_2$ monolayer, measured at low temperature ($\approx$10 K), revealing the presence of 3 resonances. The two most robust ones at $\approx$1.65 and $\approx$1.86 eV are related to the neutral A and B excitons, respectively. At lower energy side of the A resonance a weaker feature is seen, which originates from the charged exciton state. At a magnetic field of 30 T a splitting of the same magnitude is seen for the neutral A and B excitonic resonances. The charged exciton resonance becomes strongly polarized in a magnetic field, so that at 30 T its oscillator strength in the $\sigma^+$ polarization is strongly enhanced whereas the $\sigma^-$ polarization component completely disappears.

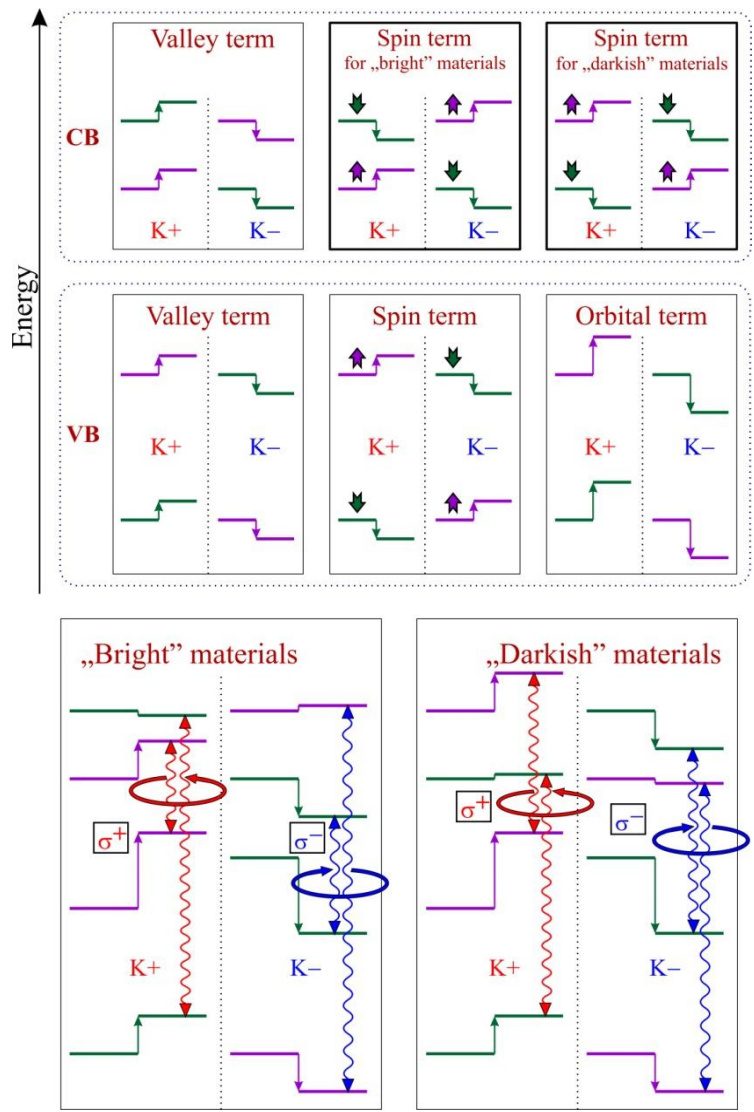

**Fig. 7.** Pictorial representation of the influence of the valley, spin and orbital terms on the energy of electronic states in an S-TMD monolayer. The arrows between the states indicate the direction, in which the particular state is shifted when the magnetic field is applied. The color code is used to indicate the spin of the states (purple for spin-up and green for spin-down states). Furthermore, in the cartoons for spin terms, additional arrow symbols are used to more clearly demonstrate the spin properties. The two possible spin configurations in the conduction band have also been highlighted in order to emphasize the origin of the two sets of optical selection rules for optically bright and "darkish" materials. In the bottom part, the two diagrams show a complete energetic landscape in a magnetic field for both types of materials. The optically active transitions between the valence and conduction states of the same spin are marked and their helicity is reflected by the following color code: red for $\sigma^+$ polarized transitions and blue for $\sigma^-$ polarized ones.

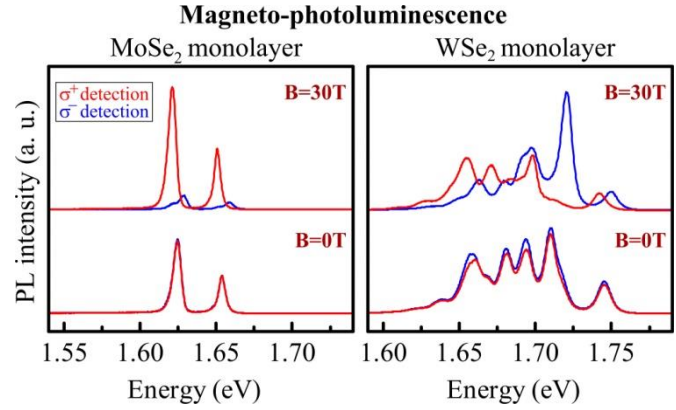

**Fig. 8.** Comparison between the low-temperature (~10 K) magneto-photoluminescence spectra of MoSe$_2$ and WSe$_2$ monolayers illustrating the differences between optically bright and "darkish" materials. For an MoSe$_2$ monolayer the neutral and charged exciton resonances are seen and their energies correspond well with the resonances from reflectivity spectra. In a magnetic field, in addition to the splitting, a strong effect of carrier relaxation results in the intensity transfer from the higher-energy ($\sigma^-$) components of both lines to the lower-energy ($\sigma^+$) ones. For the WSe$_2$ monolayer, the spectra are much more complicated due to the appearance of the low-energy multi-peak band, which additionally overlaps with the charged exciton line. The neutral exciton line splits in a magnetic field with a g-factor similar to the one of the neutral exciton in MoSe$_2$. The most peculiar finding is the huge value of g-factors of lines forming the low-energy band, reaching values as high as -14.

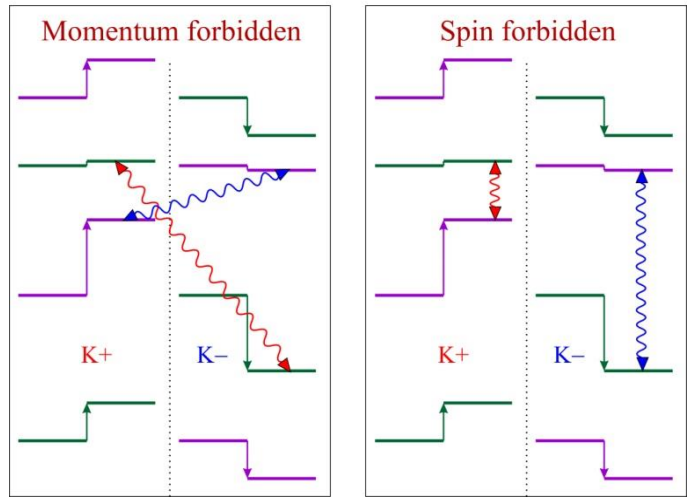

**Fig. 9.** Pictorial representation of optically forbidden transitions in a "darkish" TMD monolayer. In the left panel, the inter-valley momentum forbidden recombination is shown, which, according to the model discussed in the main text gives a value of a g-factor equal to −10. In the right panel, a spin forbidden recombination from the lower-energy conduction band is indicated, which within the proposed model implies a g-factor value of −8.

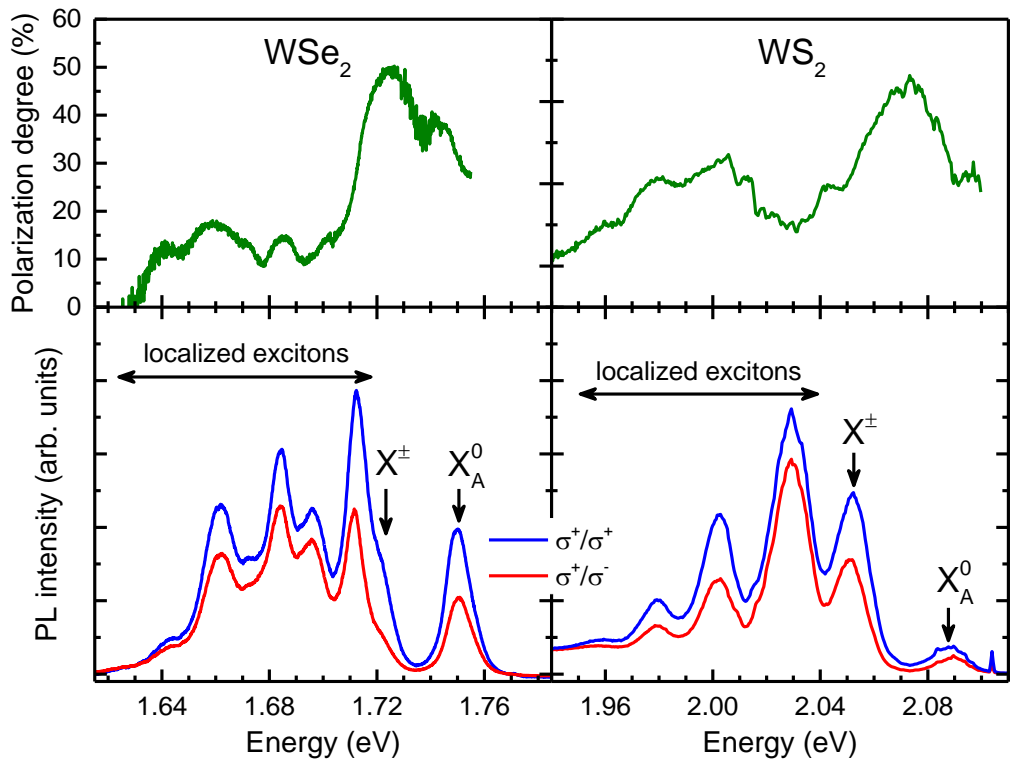

**Fig. 10.** Optical pumping experiment performed when using a σ⁺ polarized optical excitation and detecting the circularly polarized photoluminescence of monolayers of WSe$_2$ (left panel) and of WS$_2$ (right panel). The difference in the intensity in the two polarization configurations is a measure of the polarization degree. Left panel adapted from Ref. 12.

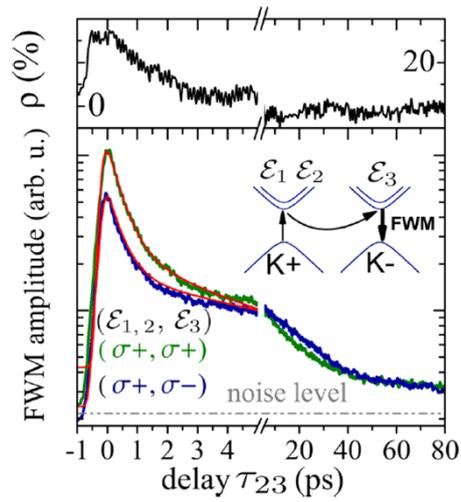

**Fig. 11.** (upper panel) Time-resolved polarization degree in an MoSe$_2$ monolayer measured close to the A exciton resonance. (lower panel) Decay of the photoluminescence intensity in both cross- and co-circular polarization configurations in the same MoSe$_2$ monolayer. Reprinted with permission from Ref. 14. Copyright 2016 American Chemical Society.

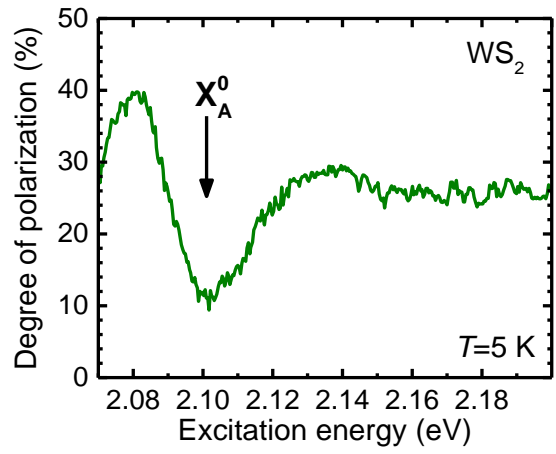

**Fig. 12.** Degree of polarization measured for a WS$_2$ monolayer, at low temperature at the charged exciton peak (X$^{\pm}$) at 2.055 eV, when sweeping the excitation laser energy across the neutral exciton resonance (X$_A^0$) which appears around 2.1 eV.

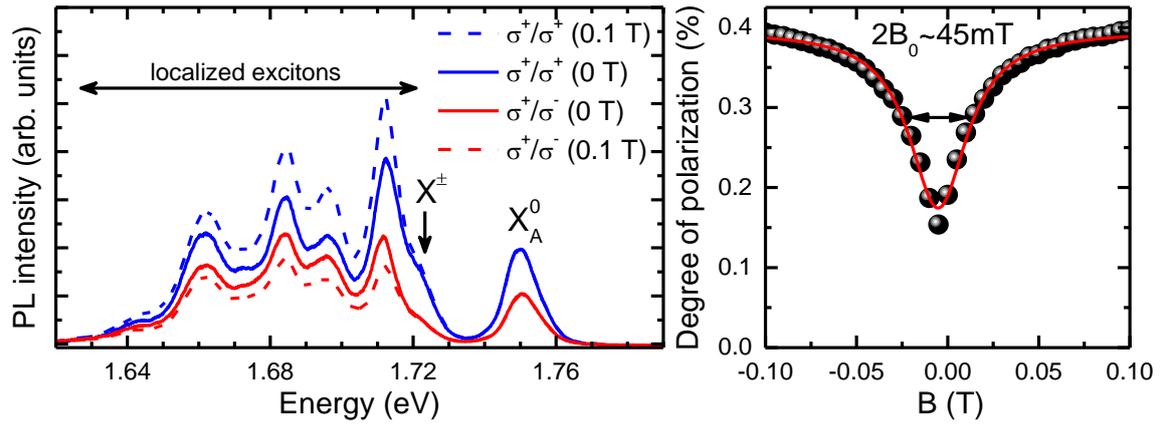

**Fig. 13.** (left panel) Polarization-resolved photoluminescence spectra of a 1L WSe$_2$ at B = 0 and when B = 100 mT is applied perpendicular to the layer (right panel). The degree of polarization measured as a function of the magnetic field applied perpendicular to the WSe$_2$ monolayer shows a strong increase of the polarization degree when a tiny magnetic field is applied. Adapted from Ref. 12.

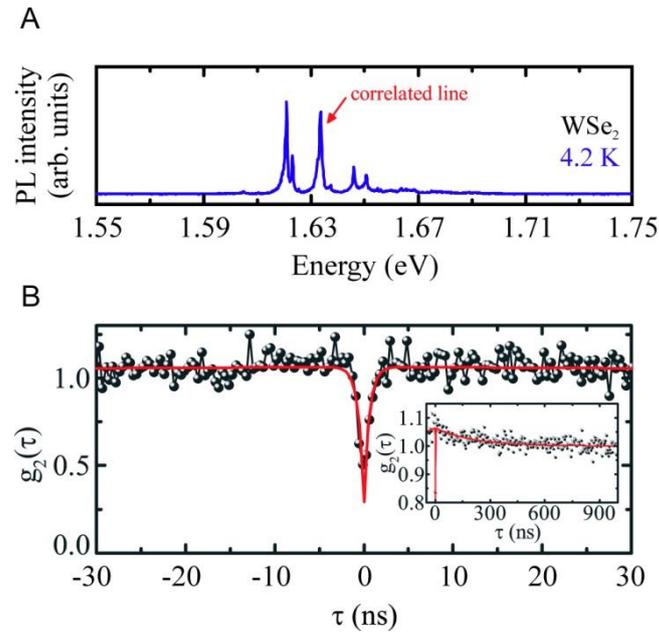

**Fig. 14. (A)** Low-temperature PL spectrum recorded at one of narrow line emitting centers located at the very edge of about 8 nm thick WSe$_2$ flake. **(B)** Photon coincidence correlation histogram displaying a second-order autocorrelation function $g_2(\tau)$, where $\tau$ denotes the time interval between the coincidence counts, measured for the narrow emission line marked in **(A)** with a red arrow. Figure adapted from Ref. 9.

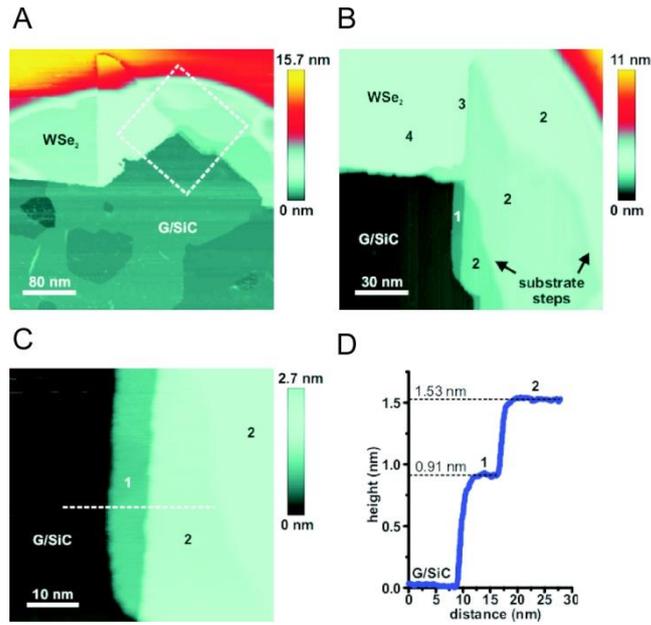

**Fig. 15.** Scanning tunneling microscopy images of the edge of a WSe$_2$ flake transferred on top of graphene (G) grown on silicon carbide (SiC) substrate, showing the existence of a nanometer-sized fragment of a WSe$_2$ monolayer at the perimeter of the flake. Panel **(B)** presents a higher-resolution image of the area marked in **(A)** with a dashed white square. Shown in **(C)** is a zoomed-in picture of the nanometer-sized monolayer with a dashed white line indicating the location of the height profile displayed in **(D)**. Figure adapted from the Supplementary Information to Ref. 9.

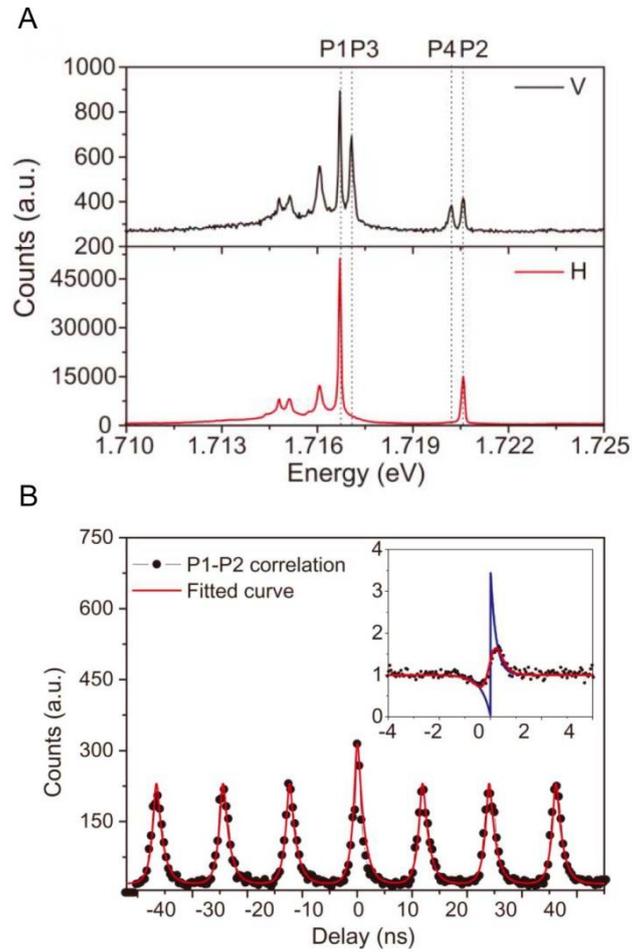

**Fig. 16. (A)** Low-temperature PL spectra of a selected narrow line emitting center hosted by a $WSe_2$ monolayer, resolved into two orthogonal linear polarizations V and H. Marked with dashed lines are two doublets of narrow emission features (P1, P3) and (P2, P4) associated with the cascaded recombination of a biexciton and exciton, respectively. **(B)** Photon cross-correlation histogram obtained for the pair of lines (P1, P2) under both pulsed (main graph) and continuous-wave (inset) excitation. Figure reprinted with permission from Ref. 120.

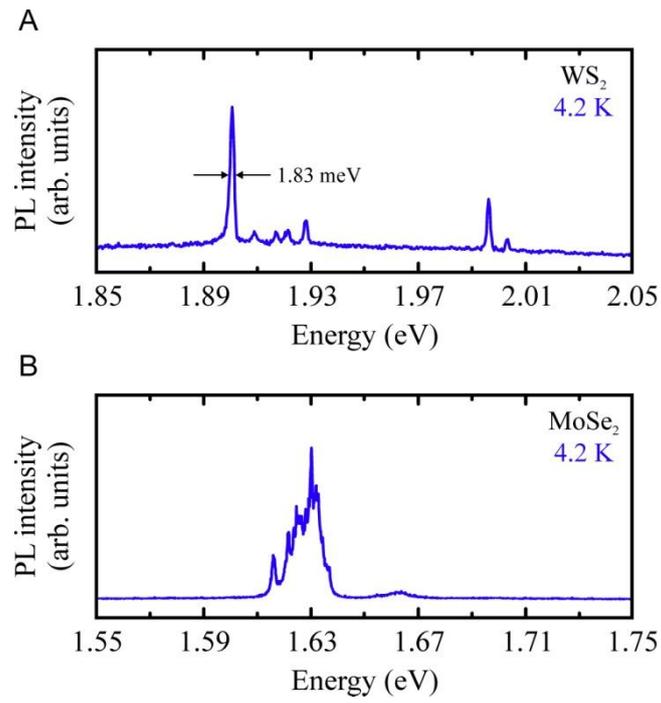

**Fig. 17.** Examples of low-temperature PL spectra measured at selected narrow line emitting centers hosted by exfoliated flakes of **(A)** WS$_2$ and **(B)** MoSe$_2$.

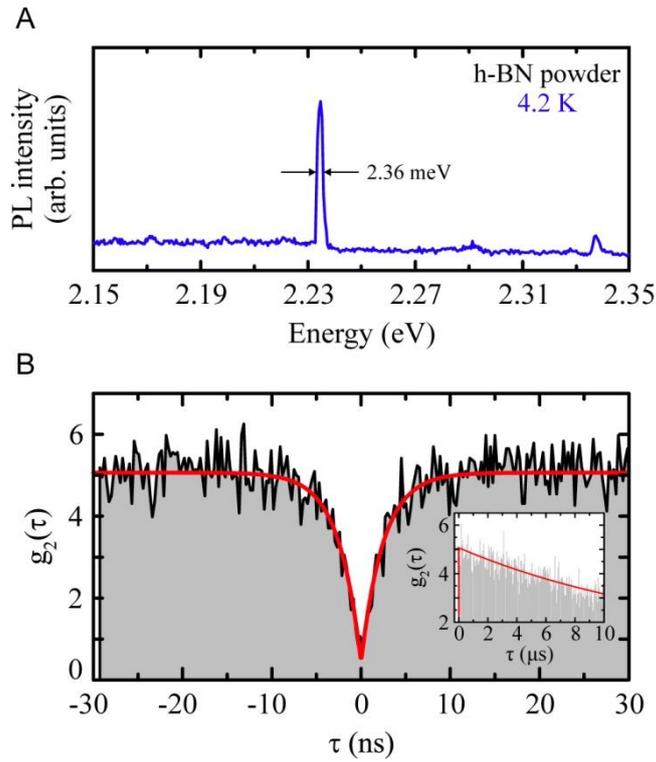

**Fig. 18. (A)** Low-temperature PL spectrum of a narrow line emitting center in h-BN powder measured at below-band-gap excitation energy. **(B)** Photon coincidence correlation histogram displaying a second-order autocorrelation function $g_2(\tau)$, where $\tau$ denotes the time interval between the coincidence counts, obtained for the narrow emission line occurring at slightly above 2.23 eV in the spectrum shown in panel **(A)**.